\documentclass[aps,prb,twocolumn,superscriptaddress,showpacs]{revtex4}
\usepackage{epsfig}
\begin{document}
\title{Spin polarization and effective mass: a numerical study\\
in disordered two dimensional systems}
\author{Richard Berkovits}
\affiliation{The Minerva Center, Department of Physics,
    Bar-Ilan University, Ramat-Gan 52900, Israel}

\date{May 9, 2004, version 1.0}
\begin{abstract}
We numerically study the magnetization of small metallic clusters.
The magnetic susceptibility is enhanced for lower electronic
densities due to the stronger influence of electron-electron
interactions. The magnetic susceptibility enhancement stems mainly
from an enhancement of the mass for commensurate fillings, while
for non-commensurate fillings its a result of an enhancement of
the Land\'e $g$ factor. The relevance to recent experimental
measurements is discussed.
\end{abstract}

\maketitle

Much attention has recently been given to the study of interacting
electrons in two-dimensional disordered systems, motivated by new
experimental observations\cite{review01}. The conductance of
dilute 2D electron systems show fascinating temperature and
magnetic field dependencies. One of the most intriguing behaviors
is the strong decrease in the critical magnetic field, $B_c$,
needed in order to fully spin polarize the system at low
densities. These low densities are characterized by a high ratio
of the intra electron interaction energy $E_c$ and the Fermi
energy $E_F$. This ratio is denoted by $r_s=E_c/E_F$. In the
weakly interacting regime ($r_s \ll 1$) the system behaves as
independent non-interacting electrons with a  magnetic
susceptibility equal to the Pauli susceptibility $\chi= g \mu_B^2
\nu$ (where $g$ is the Land\'e factor, $\mu_B$ is the Bohr
magneton and $\nu$ is the density of states at the Fermi energy).
As the electron density is lowered, measurements show an
enhancement in the susceptibility
\cite{okamoto99,vitkalov00,shashkin01,sarachik01,pudalov02,prus02}.
Although there is an ongoing debate whether these measurements
support the scenario of spontaneous spin polarization
\cite{shashkin03}, or just an enhanced magnetic
susceptibility\cite{pudalov02,prus02}, it is nevertheless generally
accepted that a large enhancement of the magnetic susceptibility
occurs over a large region of densities corresponding to $r_s>6$.

An interesting question raised recently
\cite{pudalov02,shashkin03} is whether to attribute the
enhancement of susceptibility to an increase of the $g$ factor or
to an increase of the effective mass (the density of states in a
2D system is proportional to the effective mass). By a novel
experimental method Pudalov et. al.\cite{pudalov02} were able to
measure separately $\chi$ and $m$ as function of $r_s$. Although
there are two different methods to extract $m$ which do not
exactly agree, nevertheless, it is clear that $m$ increases much
quicker as functions of $r_s$ than $g$. Recent measurements by
Shashkin et. al. \cite{shashkin03} strengthen the case for a strong
dependence of $m$ on $r_s$ while $g$ turns out to be constant.
This result is in contrast to theoretical considerations based on
the Fermi liquid picture which predict a strong enhancement in $g$
\cite{iwamoto91,kwon94,chen99}, but has some similarities to
Wigner crystallization scenarios \cite{spivak01,dolgopolov02}.

Since the development of a theoretical description of interacting
electrons in disorder systems turns out to be rather intricate, a
lot of effort has been invested in numerical studies. For example,
the influence of electron-electron interactions on the persistent
current and conductance has been extensively studied for small
metallic clusters
\cite{berkovits96,vojta98,benenti99,shepelyansky99,berkovits01,kotlyar01,selva01,berkovits02}.
In this paper we will use exact digonalization of small clusters
in order to investigate the influence of $r_s$ on $\chi$, $g$ and
$m$ of disordered systems. The magnetic susceptibility is strongly
enhanced as electron-electron interaction increases. Nevertheless,
the susceptibility shows no clear signs of divergence at any
finite value of $r_s$. While for commensurate fillings the
enhancement in $\chi$ is mainly driven by an enhancement in the
mass $m$, for non-commensurate systems the $g$
factor is strongly enhanced by interactions.

The disordered interacting cluster is represented by
the following  tight-binding Hamiltonian:
\begin{eqnarray}
{\hat H} = \sum_{k,j;\sigma} \epsilon_{k,j} n_{k,j;\sigma}
- V \sum_{k,j;\sigma} [a_{k,j+1;\sigma}^{\dag} a_{k,j;\sigma} +
\nonumber \\
a_{k+1,j;\sigma}^{\dag} a_{k,j;\sigma} + h.c. ]
+ U_{\rm H} \sum_{k,j} n_{k,j;+{1 \over 2}} n_{k,j;-{1 \over 2}}
\nonumber \\
+ U \sum_{k,j>l,p;\sigma,\sigma'} (n_{k,j;\sigma} - K)
(n_{l,p,\sigma'} - K) s / |\vec r_{k,j} - \vec r_{l,p}| ,
\label{hamil}
\end{eqnarray}
where $\vec r=(k,j)$ denotes a lattice site, $a_{k,j;\sigma}^{\dag}$
is an electron creation operator (with spin $\sigma=-{1 \over 2},
+{1 \over 2}$), the number operator is
$n_{k,j;\sigma}=a_{k,j;\sigma}^{\dag} a_{k,j;\sigma}$,
$\epsilon_{k,j}$ is the site energy, chosen randomly between $-W/2$ and
$W/2$ with uniform probability, $V=1$ is a constant hopping matrix
element, $K=n$ is a positive background charge equal
to the electronic density $n=N/M$ (where $N$ is the number of electrons
and $M$ the number of sites) and
$s$ is the lattice constant.
The electron-electron interaction is composed of the on-site
Hubbard interaction $U_{\rm H}$ between electrons of opposite spin and the
and the long range part $U$. The value of $U$ is related to the
electronic density and via $U=V\sqrt{4 \pi n} r_s$ and
we chose  $U_{\rm H} = U$ \cite{berkovits02}. Thus, the physical
content of the
density variation performed in experiment is captured by controlling
the ratio of the Fermi
energy to the interaction energy achieved
simply by changing the interaction strengths $U$.

We consider systems composed of $N=4$ and $N=6$ electrons residing
on $4 \times 3$, $4 \times 5$ and $6 \times 6$ lattices. We set
$W=8$ for the smaller lattices and $W=5$ for the larger one, so
that the single particle properties correspond to a diffusive
system., For each value of the interaction strength between
$0<r_s<30$ the six lowest eigenvalues of the sectors with $S_z=0$
and $S_z=1$ are calculated \cite{berkovits98} for a hundred
different realizations.

\begin{figure}\centering
\epsfxsize8.5cm\epsfbox{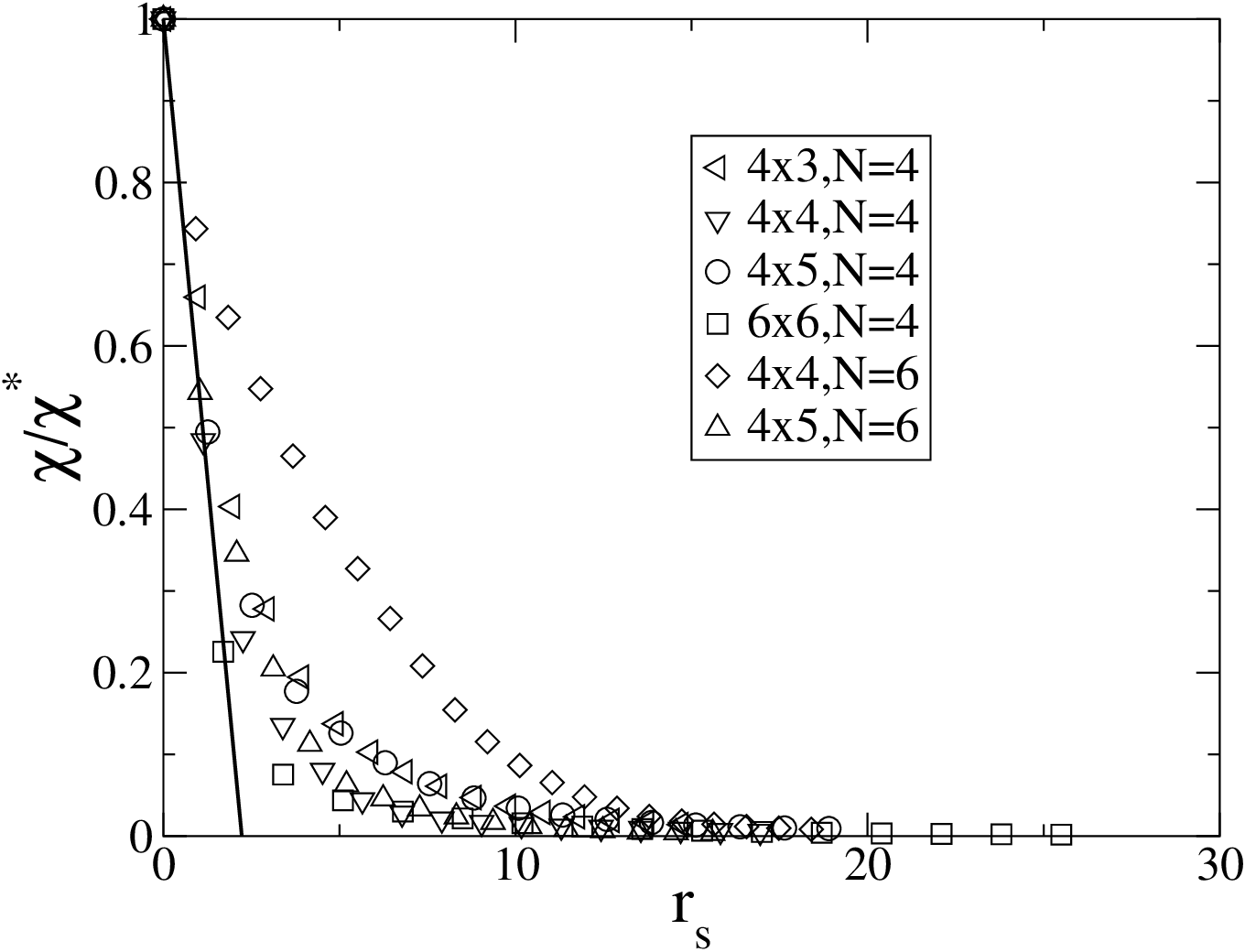} \caption{The inverse magnetic
susceptibility $\chi/\chi^*$ as function of $r_s$ for different
lattice sizes and number of electrons. The line corresponds
to $1-(\sqrt{2}/\pi)r_s$ which is the Fermi liquid correction
to the susceptibility. As the interaction
increases the susceptibility deviates from the Fermi liquid
description and depends on the lattice size and filling.} \label{fig1}
\end{figure}

\begin{figure}\centering
\epsfxsize8.5cm\epsfbox{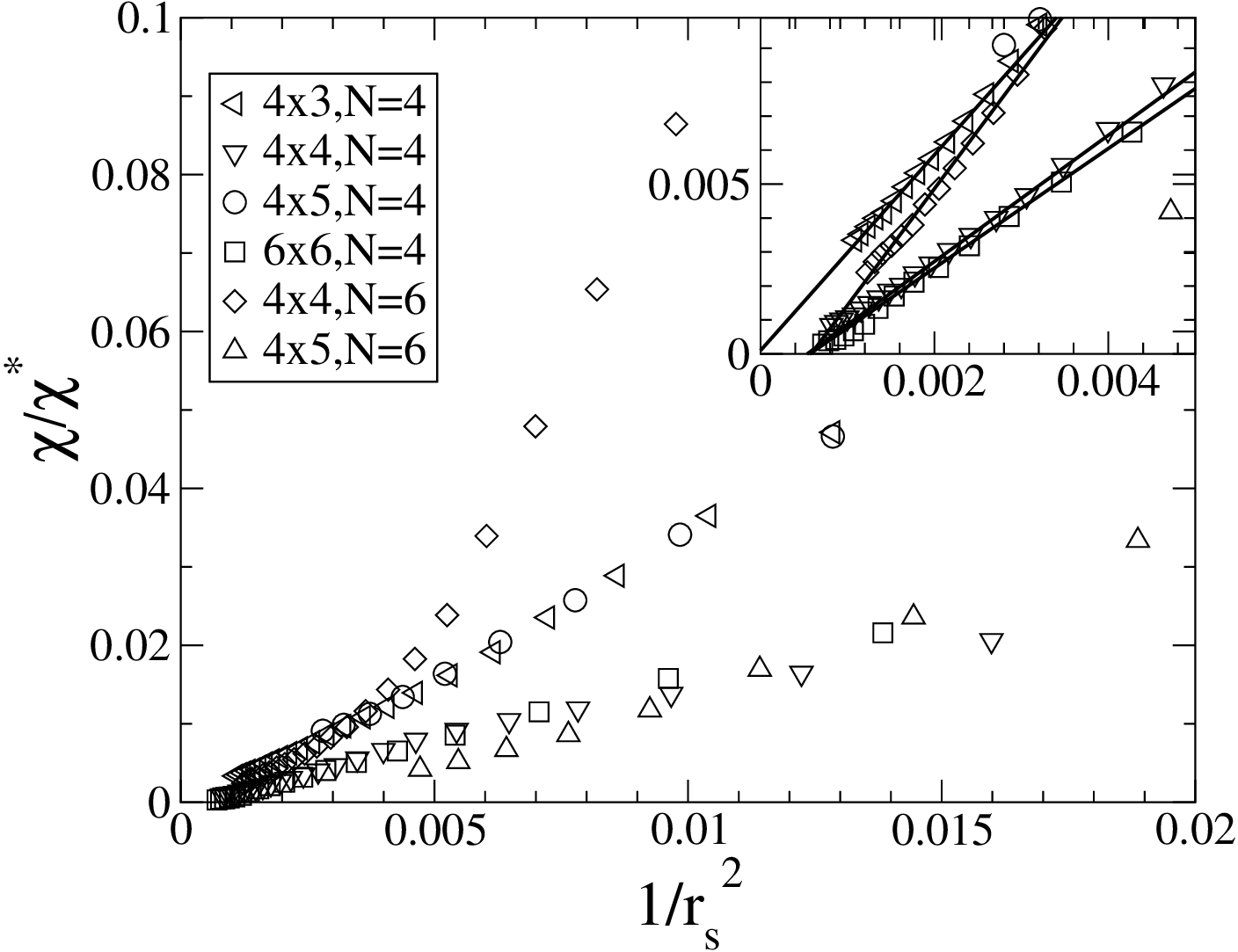} \caption{The inverse magnetic
susceptibility $\chi/\chi^*$ as function of $1/r_s^2$ (which
corresponds to the electronic density $n$) for strong interactions.
The inset focuses on the region of strong interactions. 
The lines correspond to a linear
fit of the form $1/\chi^* \propto 1/r_s^2-1/r_0^2$.} \label{fig2}
\end{figure}

Using the excitation energies it is possible to estimate the
magnetic susceptibility $\chi$ and the density of states at the
Fermi energy $\nu$. Since we are dealing with a cluster of finite
size for which the levels are discrete, it is reasonable to use a
discrete definition of the susceptibility
\begin{eqnarray}
\frac{1}{\chi^*}=\frac{\Delta B}{\Delta
M}=\frac{E_1(S_z=1)-E_1(S_z=0)}{g \mu_B^2} =\frac{1}{g^*\mu_B^2
\nu^*}, \label{sucep}
\end{eqnarray}
where $E_1(S_z)$ is the lowest eigenvalue in the corresponding
$S_z$ sector. This is based on the fact that the first change in
the magnetic moment of the system will occur when the applied
magnetic field will be strong enough to favor the $S_z=1$ state as
a ground state. That will occur once $g \mu_B B =
E_1(S_z=1)-E_1(S_z=0)$, resulting in $M = \mu_B$. A plot of
$\langle \chi \rangle / \langle \chi^* \rangle = \langle
E_1(U,S_z=1)-E_1(U,S_z=0) \rangle/\langle
E_1(U=0,S_z=1)-E_1(U=0,S_z=0) \rangle$ (where $\langle \ldots
\rangle$ denotes an average over different realizations of
disorder) is presented in Fig. \ref{fig1}. It can be seen that
$\chi^*$ is strongly enhanced as function of interaction. For
small values of $r_s$ the numerical results follow the well
known Fermi liquid predictions $\chi/\chi^*=1-(\sqrt{2}/\pi)r_s$
\cite{yarlagadda89}. 
For higher values of $r_s$,  a linear relation $1/\chi^* \propto
1/r_s^2 - 1/r_0^2$ (see Fig. \ref{fig2}) between the inverse
magnetic susceptibility and $1/r_s^2$ is observed. From this
relation one might extrapolate $1/\chi^*=0$ at $r_s=r_0$, i.e.,
full spin polarization at finite density. Noting that $1/r_s^2$
corresponds to the electronic density $n$ and that $1/\chi^*$ is
proportional to the magnetic field $B_c$ needed to fully polarize
the system, this relation may be rewritten as $B_c \propto n-n_0$
which is exactly an empirical relation proposed in Ref.
\cite{shashkin03} based on experimental observations.
Nevertheless, at large values of $r_s$ the growth in $\chi^*$
seems to tapper off, resulting in no clear evidence of divergence
in $\chi^*$ for the diffusive regime, although partial spin
polarization is seen for part of these realizations
\cite{berkovits98}.

\begin{figure}\centering
\epsfxsize8.5cm\epsfbox{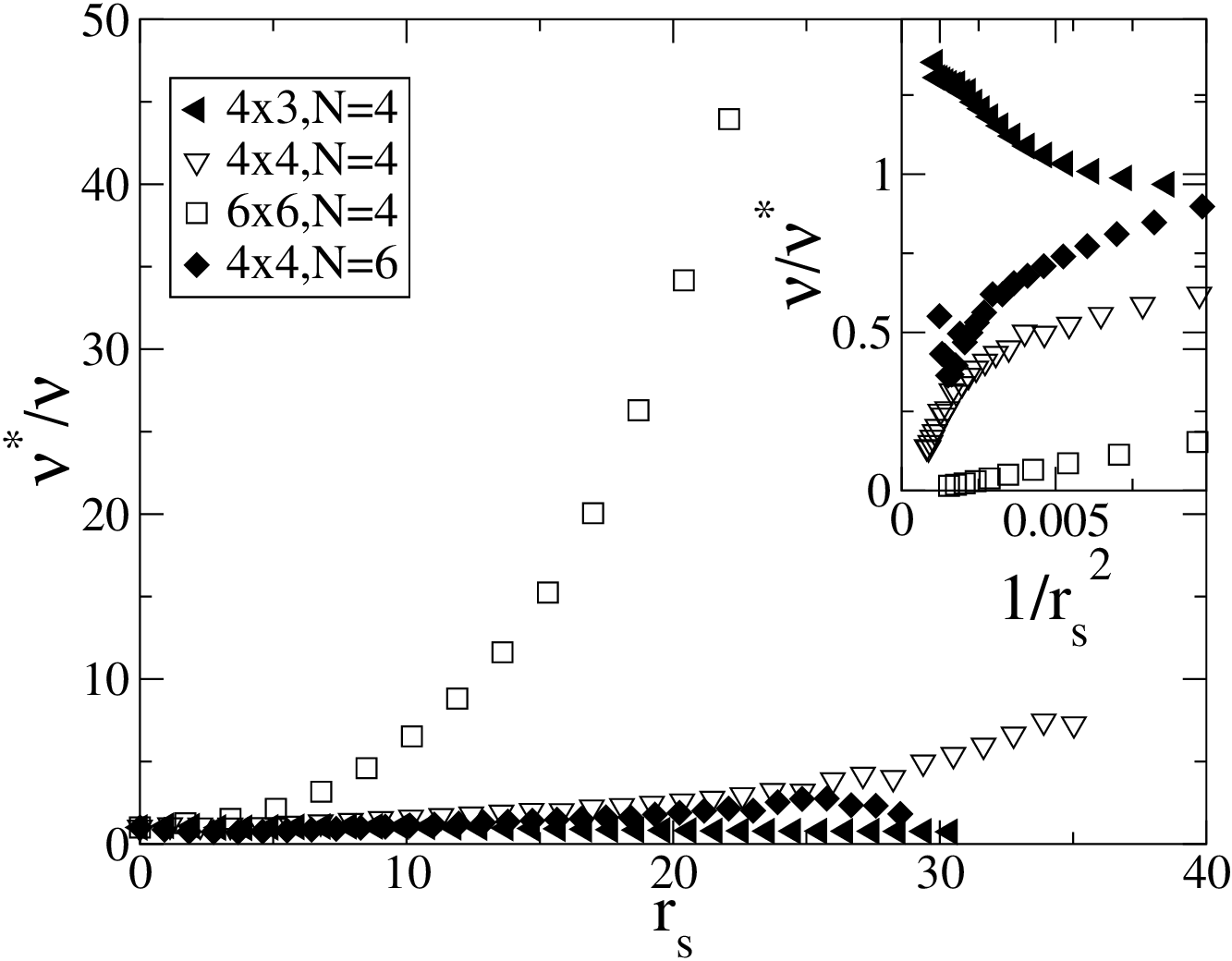} \caption{The single electron
density of states $\nu^*$  for different lattice
sizes and number of electrons. filled symbols correspond to 
non-commensurate fillings, while empty symbols to
commensurate ones. Inset: behavior of $\nu^*$ at 
strong interactions.
The density of states (i.e.,the mass) strongly increases
for the commensurate fillings, while it shows no signs
of divergence for non-commensurate fillings.} \label{fig3}
\end{figure}

\begin{figure}\centering
\epsfxsize8.5cm\epsfbox{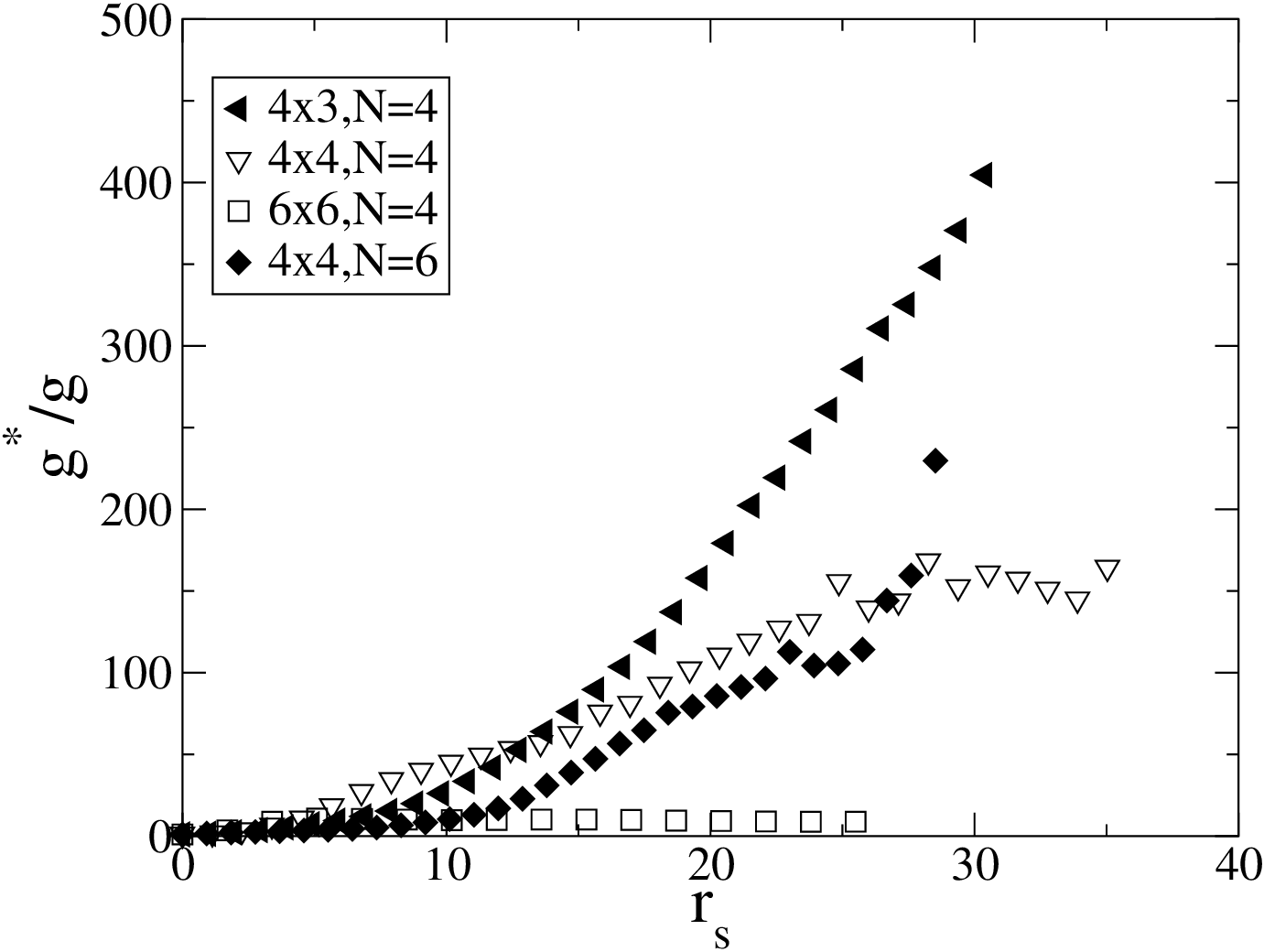} \caption{The Land\'e $g$
factor $g^*$ for different lattice sizes and number
of electrons. The $g$ factor is strongly enhanced for the
non-commensurate fillings, while it saturates for the
commensurate ones.
} \label{fig4}
\end{figure}

As in the experimental measurements, the main problem in
interpreting the enhancement in $\chi^*$ is that it might
originate from and enhancement in the $g^*$ (i.e., the result of
some exchange mechanism) or it might result from a change in the
density of states $\nu^*$ which is unrelated to the spin.
Nevertheless, it is possible to differentiate  both effects using
the following consideration. When a strong enough magnetic field
is applied on a system, the system will switch from the zero
magnetic field ground state to the first excited state with higher
spin. This process is depicted in Eq. (\ref{sucep}) and involves
both the density of states and exchange. In the Landau
quasi-particle description \cite{landau} $\chi^*/\chi=m^*g^*/mg$,
where $g/g^*=(1+F^{a}_0)$ and $m^*/m=1+F^{s}_1/2$, and $F^{s(a)}$
are the usual singlet (triplet) Fermi liquid parameters. Thus,
both the singlet and triplet Fermi liquid parameters play a role.
On the other hand, when one excites the system from the lowest
singlet state to the first excited singlet state, only the density
of states (i.e., $m$, or the singlet Fermi liquid parameter) plays
a role. The latter is accessible via the level spacing between the
two lowest excitations of the $S_z=0$ sector with total spin zero.
They can be easily identified since states in the $S_z=0$ sector
which have a total spin larger than zero, have the same energy as
their counterpart in the $S_z=1$ sector (with no magnetic field
the energy of an $S=1,S_z=0$ state is equal to the energy of a
$S=1,S_z=1$ state). Thus we identify the two lowest states in the
$S_z=0$ which are not degenerate with states in the $S_z=1$ sector:
$E_1(S=0),E_2(S=0)$ leading to
\begin{eqnarray}
\nu^*=\frac{1}{\langle E_2(S=0)-E_1(S=0) \rangle}. \label{density}
\end{eqnarray}
A plot of $\nu^*/\nu$ is shown in Fig. \ref{fig3}, while
$g^*/g=\nu^*\chi/\nu \chi^*$ is presented in Fig. \ref{fig4}. A
very different behavior between the density of states and the $g$
factor is seen. There is a strong enhancement of the density of
state as function of the interaction strength for commensurate
fillings, while the $g$ factor is strongly enhanced for the
non-commensurate fillings. Thus, for fillings which are prone to
Wigner crystallization (commensurate fillings) the mass is
enhanced , while for frustrated systems (non-commensurate systems)
the $g$ factor is enhanced. Although both the $4 \times 4$ and $6
\times 6$ lattices are at a commensurate filling for $N=4$ it is
interesting to note that the $6 \times 6$ system shows a strong
enhancement in the effective mass beginning at $r_s \sim 5$, while
the $4 \times 4$ show a strong enhancement of $m$ only much latter
($r_s \sim 30$) which is in the vicinity of the expected Wigner
crystallization transition \cite{tanatar89}. We shall comment on this
later.

Although it is rather doubtful to claim that these results on
small clusters are directly applicable for the experiment
performed on macroscopic samples, it is interesting nevertheless
to note that the behavior depicted here is in line with the recent
experimental observations \cite{pudalov02,prus02,shashkin03}. For
not to low electronic densities all groups agree that the critical
magnetic field needed to completely polarize the electronic spin
is proportional to $(n-n_0)$, which is similar to the behavior
seen in this study. Once lower densities are considered there is a
disagreement between the experimentalist. In Ref.
\cite{shashkin03} it is claimed that the linear relation holds
even for low densities and is a reliable method to find the
critical density for which full polarization occurs. On the other
hand, in Ref. \cite{prus02} doubts are raised on the applicability
of the linear relation to low densities, and whether a critical
density exist. Our numerical studies of small cluster seem to be
more in line with the latter, since we see deviations from the
linear dependence at low densities and do not see a clear
indication of full polarization at any finite density. Nevertheless,
our clusters are to small to draw firm conclusions on this point.

Another interesting point is the different behavior of the
mass and $g$ factor as function of $r_s$ between the commensurate
and non-commensurate fillings. If exchange interactions lead the
to the enhancement of the susceptibility, the $g$ factor should be
strongly enhanced, while if Wigner crystallization is responsible
for the susceptibility behavior the mass should increase.
Indeed, the dilute commensurate system ($6 \times 6$, $N=4$), which
turns into a Wigner crystal at large $r_s$, shows a strong enhancement in
the mass much prior to the expected Wigner crystallization. The
non-commensurate fillings ($4 \times 3$, $N=4$ and $4 \times 4$, $N=6$),
which do not show Wigner crystallization at large $r_s$ due to
frustration, show an enhancement in $g^*$ but not in $\nu^*$.
The commensurate system ($4 \times 4$, $N=4$) shows a rather mixed
behavior, where both $g^*$ and $\nu^*$ are enhanced. We believe 
this is due to the high filling which leads to a strong influence
of exchange correlations. Nevertheless, once $r_s$ is large enough,
$g^*$ saturates, while $\nu$ continues to grow. In the experimental
systems, which are believed to exhibit Wigner crystallization,
the large enhancement in the magnetic
susceptibility is attributed
\cite{pudalov02,shashkin03} to the enhancement of the mass.
The samples used in those
experiments are in the dilute limit and there is no reason for frustration,
i.e., they are best described by the dilute
commensurate system ($6 \times 6$, $N=4$). 
Thus, the numerical results seem
to reproduce some of the experimental behavior rather well.

In conclusion, we have numerically studied the magnetization of
small metallic clusters. The magnetic susceptibility is enhanced
by electron-electron interactions, i.e., lower electronic
densities. This enhancement stems mainly from an enhancement of
the mass in the commensurate filling. This has some similarities
to the experimental behavior recently measured.

Support from the Israel Academy of Science (Grant 276/01) is gratefully
acknowledged.

\end{document}